# Hydrostatic Pressure-enhanced correlated magnetism and Chern insulator in moiré WSe$_2$

Pengfei Jiao[1,2*], Chenghao Qian[1,3*], Ning Mao[4*], Xumin Chang[1*], Jiayong Xiao[1], Feng Liu[1], Shaozheng Wang[1], Xiaokai Wu[1,2], Di Peng[5], Cheng Xu[4], Hongliang Dong[3], Yuchen Zheng[6], Juncai Wu[6], Tong Zheng[6], Kenji Watanabe[7], Takashi Taniguchi[8], Jinfeng Jia[1,2,10,11], Xiaoxue Liu[1,2,10], Zhiwen Shi[1,2], Shiyong Wang[1,2], Guorui Chen[1,2], Tingxin Li[1,2,10], Ruidan Zhong[1,2], Yang Zhang[12,13†], Dong Qian[1,2†], Zhiqiang Chen[3†], and Shengwei Jiang[1,2†]

[1]State Key Laboratory of Micro-nano Engineering Science, Key Laboratory of Artificial Structures and Quantum Control (Ministry of Education), School of Physics and Astronomy, Shanghai Jiao Tong University, Shanghai 200240, China
[2]Tsung-Dao Lee Institute, Shanghai Jiao Tong University, Shanghai, 201210, China
[3]Center for High Pressure Science & Technology Advanced Research (HPSTAR), Shanghai, 201203, China
[4]Max Planck Institute for Chemical Physics of Solids, 01187, Dresden, Germany
[5]Shanghai Key Laboratory of Material Frontiers Research in Extreme Environments, Institute for Shanghai Advanced Research in Physical Sciences, Shanghai 201203, China
[6]Zhiyuan College, Shanghai Jiaotong University, Shanghai 200240, China
[7]Research Center for Electronic and Optical Materials, National Institute for Materials Science, 1-1 Namiki, Tsukuba 305-0044, Japan
[8]Research Center for Materials Nanoarchitectonics, National Institute for Materials Science, 1-1 Namiki, Tsukuba 305-0044, Japan
[9]Zhiyuan College, Shanghai Jiaotong University, Shanghai 200240, China
[10]Hefei National Laboratory, Hefei 230088, China
[11]Shanghai Research Center for Quantum Sciences, Shanghai 201315, China
[12]Department of Physics and Astronomy, University of Tennessee, Knoxville, TN 37996, USA
[13]Department of Physics, National University of Singapore, 117551, Singapore

**ABSTRACT**
Moiré semiconductors offer flat bands where Coulomb interactions and band topology intertwine, while the interlayer coupling plays a core role in the formation of moiré potential. However, the limited interlayer coupling strength and the lack of efficient tuning methods hinder further exploration of correlated phenomena in moiré semiconductors. We introduce a cryogenic dual-gated diamond–anvil platform using helium as a pressure medium, enabling reversible hydrostatic tuning together with magneto-optical spectroscopy in twisted bilayer WSe$_2$. Pressure enhances the moiré potential, redshifts excitons, and stabilizes Stoner ferromagnetism otherwise absent at a 3.1° twist. Simultaneously, the half-filled $C = 1$ Chern insulating state strengthens, exhibiting a reduced saturation field. Moreover, we observed a topological phase transition from Chern insulator to Mott insulator at around 2 GPa. First-principles calculations reveal that a Γ–K valence band maximum switching drives this transition by converting an Ising-like topological K-valley miniband into a spin-degenerate trivial Γ miniband. Our findings demonstrate hydrostatic pressure as a powerful, continuous control axis for correlated magnetism and topological band engineering in moiré materials.

## I. INTRODUCTION.

Moiré superlattices in Van der Waals (vdW) heterostructures are created by stacking atomically thin layers on top of one another with a twist angle or lattice mismatch. The periodic interference of the atomic lattice and the resulting moiré potential strongly modifies the original atomic bands into flat moiré minibands. Thanks to the flexibility of vdW stacking, different transition metal dichalcogenide (TMD) materials can be integrated to form hetero or homo bilayer moiré superlattices to take advantage of their different band structures. Together with the in-situ gate-tunable electric field and doping density, a huge phase space can be explored. As a result, rich quantum phenomena stemming from the interplay of strong correlation and topology have been discovered in the TMD moiré systems, including Mott insulator, generalized Wigner crystals, ferromagnetism, superconductivity, integer and fractional QAH [1–16]. The interlayer coupling plays a core role in the formation of moiré potential, leading to emergent electronic properties that are fundamentally distinct from those of the constituent layers. However, the tunability of interlayer coupling strength is limited for a specific combination of moiré materials, resulting in limited phase space for in-situ exploration. On the other hand, the interlayer coupling in moiré materials is generally weak due to its vdW nature, limiting the energy scale of the quantum phases of matter in the moiré system. Therefore, discovering an efficient method for dynamic tuning of moiré potential is highly

*These authors contributed equally to this work.
†Contact authors: yangzhang@utk.edu, dqian@sjtu.edu.cn, chenzq@hpstar.ac.cn, swjiang@sjtu.edu.cn

desirable in the study of moiré physics and applications.

In TMD moiré systems, hydrostatic pressure is a promising yet largely unexplored knob for interlayer coupling tuning. Applying pressure can effectively reduce the interlayer spacing, enhance the moiré potential[17–22], and thus modify the behavior of electrons and excitons [23–27]. Recent theory predicts that pressure can efficiently tune the flatness and quantum geometry of the single-particle bands, thus enhancing the fractional Chern insulator phase in twisted MoTe$_2$ and WSe$_2$[28,29]. Although there have been reports on hydrostatic pressure studies in TMD moiré systems, they have suffered problems of lacking the integration of high-quality samples, gate-tunability, optical access, and cryogenic-temperature conditions[18,19,21,26], preventing them from studying quantum phenomena arising from band topology and many-body interactions in TMD moiré materials, such as the correlated magnetism and Chern insulator states.

In this work, we demonstrate a cryogenically compatible diamond anvil cell (DAC) setup enabling both dual-gate control and magneto-optical spectroscopy measurement for twist bilayer WSe$_2$ moiré samples under temperature down to 1.7 K and magnetic field up to 9 T. By combining reflectance spectroscopy and magnetic circular dichroism (MCD), we observed a nonmonotonic evolution of the Stoner-type ferromagnetism and the Chern insulator with increasing pressure. Our observations are well-captured by a theoretical calculation taking into account the competition between the pressure-enhanced moiré potential and pressure-induced Γ-K valley VBM switching.

## II. DAC INTEGRATED $t$WSe$_2$

Figure 1(a) shows a schematic of the high-pressure experiment set-up. Pressure ($P$) is applied uniformly on the sample by screw-compressing the DAC chamber. Here, we use helium as the pressure-transmitting medium (PTM) to minimize the anhydrostaticity. Nonhydrostatic stress could distort the crystal lattice [30–32], which could be vital for the moiré materials due to their extreme sensitivity to external strain. Helium is considered to be the softest solid at low temperatures under high pressure, due to its large zero-point motion [33]. Even in solid helium, stress could be released through recrystallization [34], making helium the best PTM under high pressure. Previous studies demonstrated that no pressure gradients could be evidenced in helium pressure medium up to 60 GPa [35], The excellent hydrostaticity of helium PTM ensures that the $t$WSe$_2$ sample experiences uniform pressure, which is essential for studying the correlated magnetism under high pressure(see Fig. S2D).

The applied pressure is calibrated by the energy shift of the ruby R1 line [19,36,37] with an uncertainty < 5% (Fig. S2). Pre-patterned electrodes are fabricated on the DAC culets. The cubic boron nitride (c-BN) and epoxy mixture was used as an insulator layer between the BeCu gasket and electrical leads on the diamond culet to prevent short contacts [38,39]. Four Pt foils were connected to the pre-patterned electrodes, enabling electrical access to the sample (see Supplemental Material [58]). The entire DAC is maintained at a cryogenic temperature of ~1.7 K during measurements.

Figure 1(b) schematically illustrates the dual-gate device structure of $t$WSe$_2$ employed in this study. The top and bottom gate voltages ($V_t$ and $V_b$) independently control the hole filling factor ($v_h$) and the vertical electric field ($E$). $v_h = 1$ is defined as one hole per moiré unit cell, corresponding to a half-filled moiré valence band. By combining reflective MCD and reflectance contrast (RC) data, we could identify the filling factor ($v_h = 1/3, 1$) and infer a twist angle of ~3.1° (see Supplemental Material [58]). The $t$WSe$_2$ active layer is fully encapsulated between two sets of graphite/$h$-BN layers, which serve as top and bottom gate electrodes and dielectric, respectively. The encapsulation also prevents the perturbation of PTM and DAC surface, helps maintain ultra-clean interfaces, and ensures the high device quality in the whole pressure range.

Figure 1(c) illustrates the honeycomb moiré lattice with two sublattices centered at the MX and XM (M=W; X=Se) stacking sites. A schematic illustration of the band structures (valence band only) for $t$WSe$_2$ under ambient and high pressure is shown in Fig. 1(d). In the vicinity of the K point of the Brillouin zone, the band structure of each WSe$_2$ layer can be regarded as a single spin-polarized parabolic band. Interlayer coupling caused a strong hybridization of the bands where they intersect, leading to moiré minibands. Without pressure, the moiré miniband from K valley (KMB) is lower in energy compared with the moiré miniband from Γ valley (ΓMB) [40–42]. It has been demonstrated that applying pressure can reduce the interlayer distance, enhance the interlayer coupling, which effectively enhances the moiré potential [17–22]. On the other hand, the relative energetic positions between the K and the Γ valleys of TMD's valence band are very sensitive to the interlayer coupling due to their different orbital characters. A notable example is that the introduction of interlayer coupling turns the direct K-K gap in monolayer TMD into an indirect Γ-K gap in natural bilayer TMD. Recent theory predicts that applying uniaxial pressure to a twisted bilayer TMD switches the VBM from K valley to Γ valley, thus switching the moiré lattice model [40–42].

*These authors contributed equally to this work.
†Contact authors: yangzhang@utk.edu, dqian@sjtu.edu.cn, chenzq@hpstar.ac.cn, swjiang@sjtu.edu.cn

Figure 1(e) shows the optical reflectance contrast spectrum as a function of $v_h$ at $E = 0$ (see

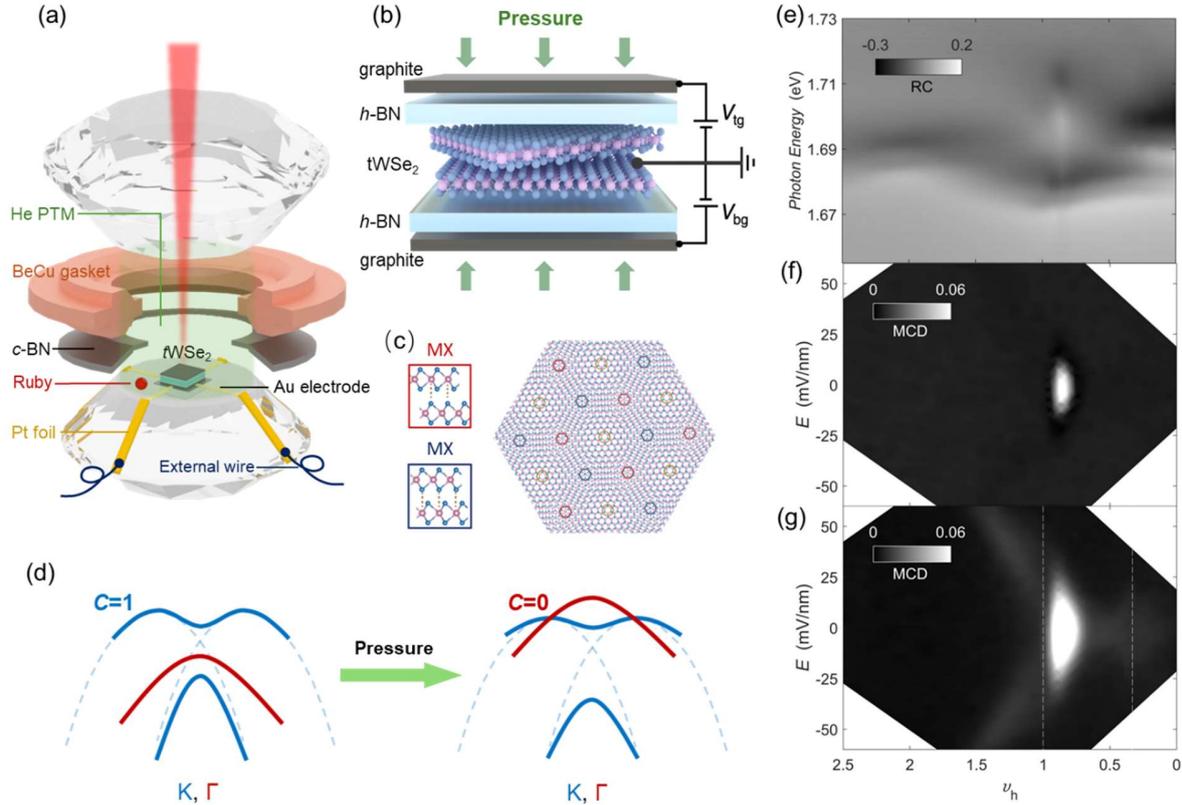

FIG 1. DAC integrated $t$WSe$_2$ device. (a), Schematic illustration of the high-pressure experimental setup using a DAC device. The $t$WSe$_2$ sample is directly fabricated on one of the diamond anvils, and a ruby microsphere is placed nearby for pressure calibration. The pressure-transmitting medium is helium. Pt foils are manually mounted as feedthrough wires, allowing electrical access. The DAC is maintained at a temperature of 1.7 K during measurements. (b), Schematic side view of the $t$WSe$_2$ device under pressure. The dual-gate structure allows an independent tuning of the doping and electric field. (c), The schematic moiré superlattices of R-stacked $t$WSe$_2$. High-symmetry stackings are highlighted by circles. The red circles correspond to MX sublattice sites, where the W atoms in the top layer are aligned with the Se atoms in the bottom layer. Blue circles are the corresponding XM sublattice sites. MX and XM are two degenerate energy minima at $E = 0$, forming a honeycomb moiré superlattice. (d), Schematic representation of the $t$WSe$_2$ hybridized band structure under ambient pressure (left) and high pressure (right). The blue and red denote bands from K and Γ valley, respectively. The stronger interlayer coupling under high pressure leads to a larger hybridization gap. The valley order also swaps from K to Γ under high pressure. (e), RC spectrum as a function of $v_h$ at $E = 0$. (f),(g), MCD maps as a function of $v_h$ and $E$ under 0.01 T (f) and 0.1 T (g). $v_h = 1$ and $1/3$ are indicated by the vertical dashed lines.

Supplemental Material for electrostatics [58]), measured under ambient pressure before closing the DAC assembly (see Supplemental Material [58]). We observed that the attractive polaron (AP) resonance exhibits modulation and oscillator-strength transfer at integer filling factors $v_h = 1$ and 2 (see Supplemental Material for details [58]), signifying the effect of moiré potential [43,44]. Figures 1(f) and 1(g) show the corresponding MCD intensity map plot as a function of $v_h$ and $E$, under 0.01 T and 0.1 T. A strong MCD signal with a clover-shaped pattern is observed under B=0.1 T, indicating the parameter space with large magnetic susceptibility ($\chi_{MCD}$). The $\chi_{MCD}$ also show a feature of enhancement at $v_h = 1/3$ (Fig. S4). When the $B$-field is reduced to 0.01 T, we find the region with clear MCD signal shrinks to a small hotspot, centering at $v_h \sim 0.85$. The above observation is consistent with the earlier report that the regions with enhanced magnetic susceptibility are correlated with the electronic DOS, which is substantially enhanced at the van Hove singularity (vHS) [43,45], because the magnetic susceptibility of a Landau Fermi liquid is proportional to the electronic DOS [46]. In the correlated insulator state at $v_h = 1$, the Fermi liquid picture breaks down, and the susceptibility arises from the local magnetic moments. Indeed, we observed a weak strip-like region with suppressed $\chi_{MCD}$ at $v_h = 1$ in Figure 1(g).

*These authors contributed equally to this work.
†Contact authors: yangzhang@utk.edu, dqian@sjtu.edu.cn, chenzq@hpstar.ac.cn, swjiang@sjtu.edu.cn

Although substantially enhanced $\chi_{MCD}$ is observed, the absence of remanent magnetization and hysteresis (Fig. 3(a) and 3(e)) signifies the absence of ferromagnetic (FM) order at the base temperature of 1.7 K. On the other hand, FM order is observed in the 2.7° $t$WSe$_2$ sample in the earlier report [43]. The discrepancy could be qualitatively understood by the twist angle-dependent interaction parameter $r_s = U/W$ and Stoner criterion. The Stoner criterion [47], $UD_F > 1$, expressed in terms of the strength of Coulomb repulsion $U$ and the single-particle DOS at the Fermi level $D_F$, provides a qualitative threshold for magnetism. For a simple parabolic 2D band, we have $U \sim 1/a_m$, $W_m \sim 1/a_m^2$ and $D_F = 1/W_m$, where $a_m$ is the moiré lattice constant, and $W_m$ is the width of the moiré miniband. The left side of the Stoner criterion can be expressed as $\frac{U}{W_m} = r_s \sim a_m$. Therefore, the Stoner criterion is better satisfied at a smaller twist angle with larger $a_m$. This picture is further supported by the recent study, which shows that the FM vanished while superconductivity emerges in $t$WSe$_2$ with larger twist angles of 3.5° [15].

### III. PRESSURE-ENHANCED MOIRÉ POTENTIAL

Figure 2(a) and 2(b) show the pressure evolution of the RC spectrum as a function of $v_h$ at $E = 0$. It is worth noting that after the DAC assembly is closed with the top diamond, the RC magnitude is significantly reduced, likely due to the strong reflection from the diamond-vacuum and diamond-PTM interface. Nevertheless, we observed the modulation of AP resonance at $v_h = 1$ and 2 under pressure up to 2.1 GPa, signifying the robustness of the moiré potential under pressure [43,44]. Additionally, we observe a continuous red shift of the exciton resonances with increasing pressure, as summarized in Figure 2(c). Earlier studies have shown that the enhancement of moiré potential reduces the exciton resonance energy [17,48,49]. Therefore, the red shift of exciton resonances signifies that the moiré potential in $t$WSe$_2$ is indeed enhanced by the hydrostatic pressure. The above observation is consistent with previous reports that applying pressure reduces the interlayer distance, enhances the interlayer coupling, which effectively enhances the moiré potential[17–22].

When the pressure increases to 2.1 Gpa (Fig. 2(b)), we see a resolvable change in the RC spectrum. While the modulation feature is still evident, the AP resonance is overall broader and less pronounced. We attribute this evolution to the pressure-driven Γ–K crossover, in which the system transitions from a direct gap (K–K) with relatively large optical oscillator strength to an indirect gap with reduced oscillator strength, as will be discussion later. The pressure is then released back to 1.1 Gpa to check if there is any irreversible change happening after high pressure. We find the modulation feature in AP resonance recovers at 1.1 Gpa, evidencing the reversible pressure-engineering process. The good reversibility is also consistent with the excellent hydrostaticity of the helium PTM.

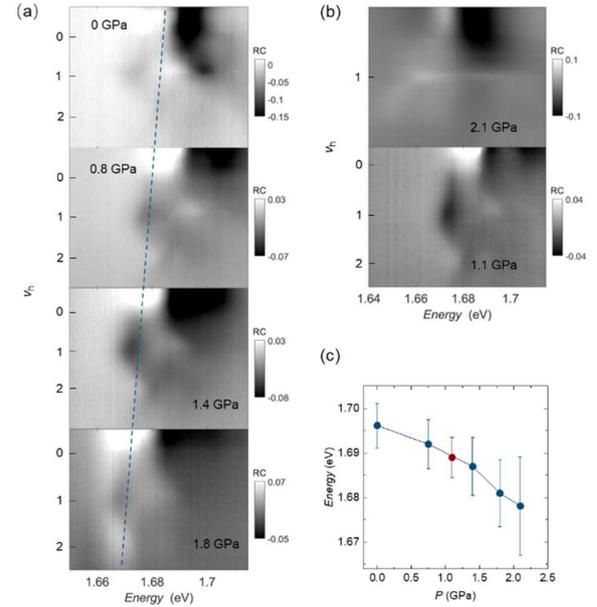

FIG. 2. Pressure dependence of moiré exciton resonance. (a), (b), RC spectrum as a function of $v_h$ at $E = 0$, under representative pressures. The dashed line serves as the guide-to-the-eye, highlighting the red shift of the moiré exciton resonance under increasing pressure (a). The AP resonance is recovered after pressure is released back from 2.1 GPa to 1.1 GPa (b), evidencing the reversible pressure-engineering process. (c), Resonance energy of the neutral moiré exciton as a function of pressure, extracted from (a) and (b). The red data point represents the measurement taken at 1.1 GPa during the decompression cycle (after the sample was released from 2.1 GPa).

*These authors contributed equally to this work.
†Contact authors: yangzhang@utk.edu, dqian@sjtu.edu.cn, chenzq@hpstar.ac.cn, swjiang@sjtu.edu.cn

## IV. PRESSURE TUNING OF FERROMAGNETISM AND CHERN INSULATOR

To investigate the pressure effect on the correlated magnetism in $t$WSe$_2$, we performed MCD measurements under various pressures (see Supplemental Material [58]). Figure 3(a)-(d) shows the MCD intensity map plot as a function of $v_h$ and $E$ under representative pressures, under 1 T and in the absence of magnetic fields. Figure 3(e) shows the corresponding MCD-$B$ curves at $E = 0$ and $v_h \sim 0.85$ (with the strongest magnetic response). Fig. 3(a) shows the MCD map under ambient pressure. While the MCD signal with a clover-shaped pattern is observed under $B = 1$ T, no MCD signal can be resolved in the zero-field MCD map. The absence of remanent magnetization and hysteresis (Fig. 3(e)) signifies the absence of ferromagnetic (FM) order.

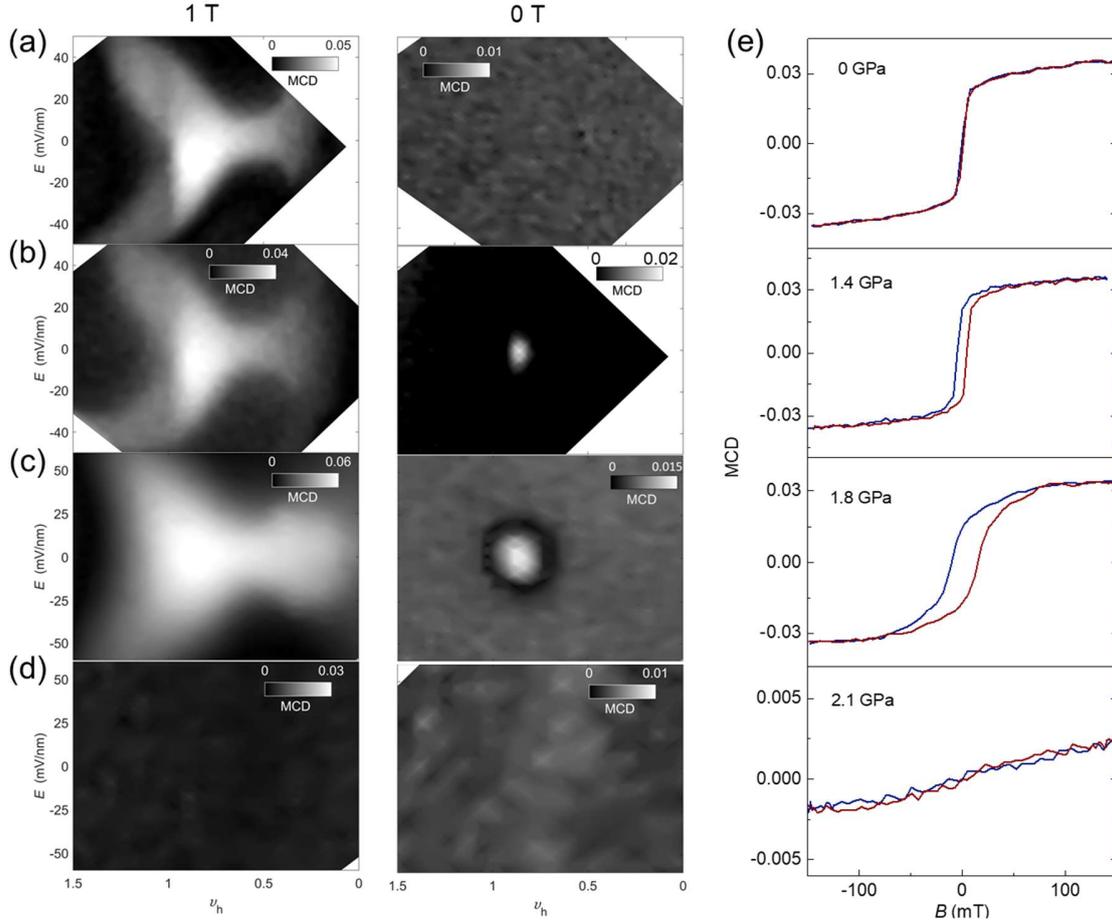

FIG. 3. Pressure dependence of magnetic properties. (a)-(d), MCD maps as a function of $v_h$ and $E$ under representative pressures and 1 T (left panel) and 0 T (right panel), respectively. The measurement is performed after closing the sample space with the pre-indented gasket and aligning the other piece of diamond. (e), MCD as a function of $B$ at $v_h \sim 0.85$, under representative pressures. Measurements are performed at 1.7 K.

Fig. 3(b) and 3(c) show the MCD map under $P = 1.4$ and 1.8 Gpa. Remarkably, the remanent MCD signal appears at zero magnetic field. The MCD-$B$ curves exhibit clear magnetic hysteresis with finite coercive fields, signifying the emergence of FM order.

The stabilization of FM order under pressure could be understood considering the pressure-enhanced interlayer coupling and the Stoner criterion. The bare Coulomb potential $V(q)$ (giving the Hubbard $U$ when projected onto onsite Wannier basis) of electrons in a material will be screened by the polarization of other electrons, resulting in the screened Coulomb interaction [50,51]:

*These authors contributed equally to this work.
†Contact authors: yangzhang@utk.edu, dqian@sjtu.edu.cn, chenzq@hpstar.ac.cn, swjiang@sjtu.edu.cn

$$V_s(q,\omega) = \frac{V(q)}{\varepsilon(q,\omega)}$$

where $\varepsilon(q,\omega)$ and $V_s(q,\omega)$ represent the dielectric function and screened Coulomb interaction at wavevector $q$ and frequency $\omega$. $\varepsilon(q,\omega) = 1 - V(q)\chi_0(q,\omega)$, where $\chi_0(q,\omega)$ is the Lindhard function. In the static limit ($\omega \to 0$), the dielectric function becomes [51]:

$$\varepsilon(q) = 1 + V_q \frac{4}{\Omega} \sum_k \sum_{cv} \frac{|\langle \psi_{vk}|e^{-i\,\cdot\,r}|\psi_{ck+q}\rangle|^2}{E_{ck+q} - E_{vk}}$$

where $\Omega$ is the area of the unit-cell, $\psi_{ck+q}$ and $\psi_{vk}$ are the wavefunction, and $E_{ck+q}$ and $E_{vk}$ are the energies of the conduction and valence bands at momenta of $k+q$ and $k$, respectively. In $t$WSe$_2$, we focus only on the hole doping physics and take the small $q$ limit, since the momentum scale of moiré superlattice is very small. Therefore, the indices $v/c$ refer to the first and second valence bands. Consequently, we find that $\varepsilon(0) \sim 1/E_g$, where $E_g$ is the interlayer-hybridization-induced band separation between the first and second bands. Increasing pressure enhances the orbital overlap between the top and bottom layers, thereby enlarging $E_g$ (Fig. 4(j)) and reduces dielectric screening $\varepsilon$. At the same time, the bandwidth of the partially filled first K-valley band decreases with increasing pressure, $\approx 20\%$ reduction from 0 to 1.8 Gpa (Fig. 4(k)).

The reduction in bandwidth enhances the bare Coulomb potential $V(q)$, as well as the density of states $D_F$. As a result of the enhanced $V_s$ together with the enhanced $D_F$, Stoner criterion is more readily satisfied, resulting in the stabilization of ferromagnetism.

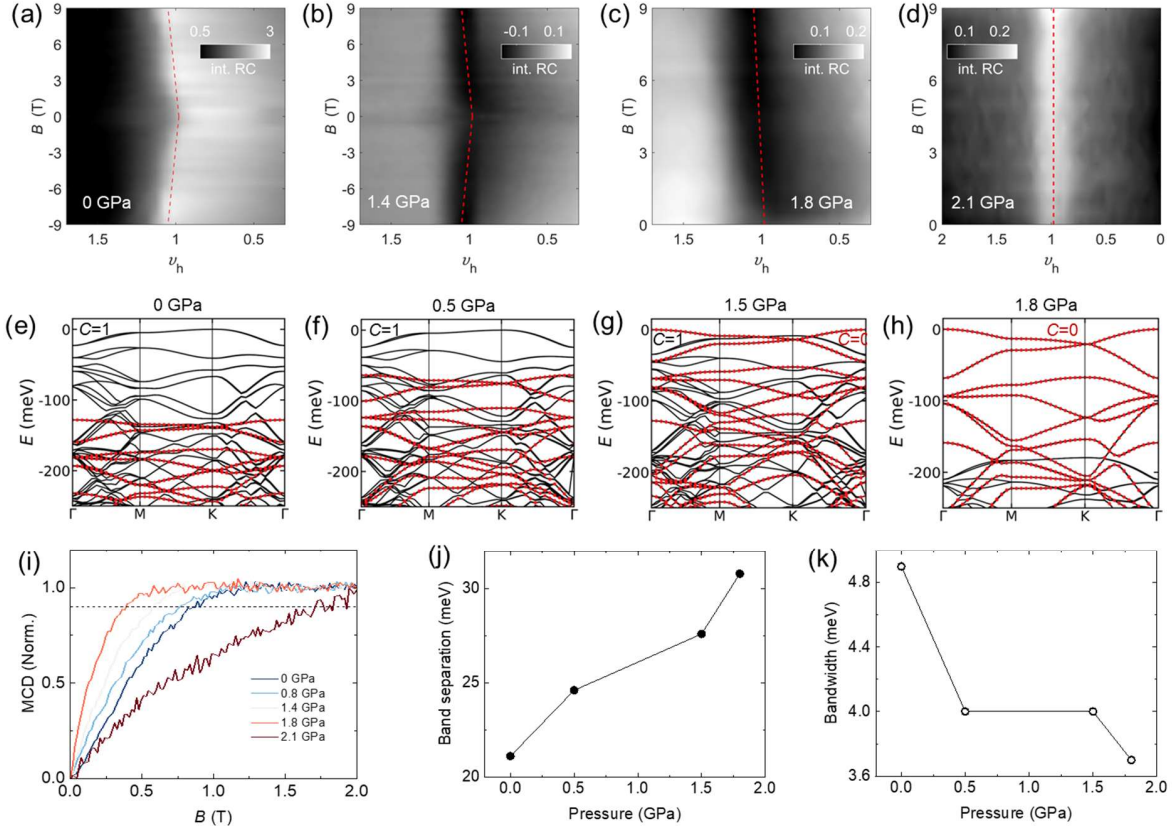

FIG. 4. Pressure dependence of the Chern insulator states and theoretical calculation. (a)-(d), Spectrally integrated RC intensity versus $B$ and $v_h$ at representative pressures. Dashed lines represent the expected dispersions based on the Streda formula for $v_h = 1$. (e)-(h). First-principles band structures of $t$WSe$_2$ with a twist angle of 3.15° at ambient pressure (e), under 0.5 GPa (f), 1.5 GPa (g), and 1.8 GPa (h). The red bands highlight the Γ-valley states. At ambient pressure, the topmost valence band originates from the K-valley and carries Chern number of 1. Upon applying pressure, the Γ-valley bands shift upward and become the highest valence states, resulting in the trivial first moiré mini-band with $C = 0$. (i), Magnetic-field dependence of the spectrally integrated MCD for $v_h = 1$ under different

*These authors contributed equally to this work.
†Contact authors: yangzhang@utk.edu, dqian@sjtu.edu.cn, chenzq@hpstar.ac.cn, swjiang@sjtu.edu.cn

pressures. Dashed lines indicate the saturation field, which is defined as the field where MCD reaches 90% of its saturated value at high fields. (j), Evolution of band separation with pressure from first-principles calculation. The band separation is defined as the direct energy gap between the first and second K-valley bands at the M point. As pressure increases, the band separation becomes larger, indicating reduced Coulomb screening. (k), Evolution of bandwidth with pressure from first-principles calculation. The bandwidth is defined as the energy spread along the Γ–M path, considering only the partially filled first K-valley band. With increasing pressure, the bandwidth decreases, suggesting enhanced effective Coulomb interactions in the system.

When pressure is further increased to 2.1 Gpa, the MCD response becomes very weak (Fig. 3(d)). MCD-$B$ curve becomes almost linear, and no evident $\chi_{MCD}$ enhancement is observed at small $B$-field (Fig. 3(e)). The nonmonotonic pressure dependence of the Stoner-type ferromagnetism is also supported by the nonmonotonic pressure dependence of the Weiss temperature $T_w$ (Fig. S5).

To further understand the observed nonmonotonic pressure dependence of the Stoner-type ferromagnetism, we performed first-principles calculations for the electronic band structure of $t$WSe$_2$ under out-of-plane pressures (Fig. 4(e)-4(h)) (see Supplemental Material for details [58]).

Under ambient pressure (Fig. 4(e)), the VBM is at the K/K' points. Additionally, there exists a local VBM at the Γ valley. The K/K' valleys come from predominantly the in-plane orbitals $d_{xy} \pm id_{x^2-y^2}$ of W, the band has large spin-orbit splitting (hundreds of meV) originating from the orbital angular momentum, showing effective Ising spin anisotropy. The orbital content of Γ valley is predominantly the out-of-plane $d_{z^2}$ of W. The Γ bands are spin-degenerate, showing negligible spin anisotropy [52]. The energy difference between the Γ and K valley ($\Delta_{\Gamma K}$) of about 90 meV has been observed in the photoemission spectroscopy study of $t$WSe$_2$ [53,54], in reasonable agreement with our calculation of ~130 meV.

With increasing pressure, the interlayer distance decreases, and the energy of the Γ valley increases significantly. In contrast, the energy of the K valley is less affected. Physically, this effect stems from the fact that the interlayer hybridization of the out-of-plane $d_{z^2}$ orbitals are much larger than the in-plane $d_{xy} \pm id_{x^2-y^2}$ [40,41].

As pressure increases to ~1.5 GPa, first-principles calculations show that a Γ-K valley changeover occurs. The VBM changes from K to Γ in calculation (Fig. 4(g)), and the doped holes will transfer from a spin-orbit-split KMB to a spin-degenerate ΓMB. Considering spin anisotropy, the K-Γ valley switching is accompanied by the switching of the spin model: from effective 2D Ising to effective 2D Heisenberg spin system. While finite-temperature magnetic order is allowed in a 2D Ising spin system, enhanced thermal fluctuation will suppress finite-T order per Mermin–Wagner for 2D Heisenberg-like limit [55]. Therefore, this valley-and-anisotropy switching will suppress finite-temperature long-range Stoner-type ferromagnetic order. Our observation of suppressed FM order and higher pressure is likely aligned with this scenario.

To study the pressure effect on the topological properties in $t$WSe$_2$, we measure the magnetic field- and filling-dependence of the RC spectra at various pressures. Chern numbers ($C$) of the correlated states can be extracted based on the Streda formula, by tracking their density shift as a function of magnetic field [10,43,44,56,57].

Figure 4(a)-4(d) displays the pressure evolution of the correlated insulators at $v_h$ =1 under magnetic field, i.e., the spectrally integrated AP intensity as a function of $v_h$ and $B$. In the zero-pressure limit (Fig. 4(a)), the insulating state manifests as a dispersive shape. At intermediate pressures (Fig. 4(b)–(c)), this feature evolves into a distinct dip, while at 2.1 GPa (Fig. 4(d)), it appears as a local peak. This variation in spectral lineshape is a well-known effect in thin-film optics, attributed to the pressure-dependent changes in the optical interference conditions within the DAC cavity, rather than a change in the intrinsic nature of the excitonic resonance (see Supplemental Material for details [58]). Despite these variations in contrast, we observe that the spectral feature disperses linearly with $B$ at all pressures up to 1.8 GPa. The dispersion agrees well with $C = 1$ according to the Streda formula $n_M \frac{dv_h}{dB} = C \frac{e}{h}$, as illustrated by the dashed lines. The observed Chern insulator state at $v_h$=1 is consistent with the earlier reports [43,44]. Moreover, we found the saturation field ($B_s$) at $v_h$ =1 decreases with increasing pressure, from ∼0.8 T under ambient pressure to ∼0.4 T under 1.8 GPa (Fig. 4(i)), indicating that the Chern insulator state is enhanced by the pressure. This is likely due to the pressure-enhanced moiré potential and thus the electron correlation. When further increasing pressure to 2.1 GPa, the dispersion disappeared (Fig. 4(d)), suggesting that the occupied miniband becomes topologically trivial.

The above observation is consistent with the expectation that the valley-switching will simultaneously transfer the holes from the Berry-curvature-rich topological K miniband ($C = 1$) to a trivial ($C = 0$) half-filled state, which is supported by our band topology calculation (Fig. 4(e)-4(h)). To determine the topological character of the first K-valley and Γ-valley bands, we evaluate the $C_3$ rotation eigenvalues at the $C_3$-invariant momenta (Γ, K, and K'). For the topmost K-valley band, the eigenvalues are approximately $π/3$, $π$, and $π$, whose product corresponds to a Chern number $C = 1$. For the topmost Γ-valley band, the eigenvalues at all three high-symmetry points become approximately $-π/3$, giving $C = 0$ (see Supplemental Material [58]). Therefore, we attribute the vanishing of the Chern insulator state to the pressure-induced Γ-K valley VBM switching, which drives a topological phase transition from a nontrivial to a trivial phase. To the best of our knowledge, this is the first experimental demonstration of pressure-induced topological phase transition from Chern insulator to Mott insulator in moiré materials systems.

*These authors contributed equally to this work.
†Contact authors: yangzhang@utk.edu, dqian@sjtu.edu.cn, chenzq@hpstar.ac.cn, swjiang@sjtu.edu.cn

## V. CONCLUSIONS

In summary, we have developed a cryogenically compatible DAC setup enabling both dual-gate control and magneto-optical spectroscopy measurement for $t$WSe$_2$ moiré samples. By using helium as PTM, we achieved excellent hydrostaticity. The moiré potential in $t$WSe$_2$ can be effectively enhanced by pressure, manifested by the red shift of the exciton resonances with increasing pressure. The enhanced interlayer coupling leads to stronger electron correlation, resulting in the stabilization of correlated ferromagnetism and Chern insulator state. When large pressure above ~2 GPa is applied, we observed the suppression of the ferromagnetism, which likely stems from the Γ-K valley transfer. The pressure-induced Γ-K valley transfer is further supported by the suppression of the Chern insulator phase under large pressure.

Our work establishes hydrostatic pressure as a continuous dimension orthogonal to twist, gating, displacement field; together they create a higher-dimensional parameter space for optimizing flatness, Berry curvature, and interaction scales. Our study also paves the way for pressure-engineering exotic quantum phases of matter arising from band topology and many-body interactions in moiré materials, such as the extension of fractional quantum anomalous Hall insulator [28,29], and potential interplay with emerging superconductivity at larger angles [14,15]. Although superconductivity is absent in our 3.1° sample, the pressure tuning of superconductivity in WSe$_2$ moiré system is a very interesting direction to explore, as the pressure usually enhances the superconductivity. However, it requires more complex device structure, more electrodes/wiring and millikelvin dilution fridge integrated high pressure setup, which is beyond the scope of this study.

## ACKNOWLEDGMENTS

Research was primarily supported by the National Key R&D Program of China (Grants Nos. 2021YFA1401400, 2021YFA1400100, 2022YFA1402400, 2022YFA1405400, 2022YFA1402702, 2022YFA1402404, 2019YFA0308600, 2022YFA1402401, 2020YFA0309000), the National Natural Science Foundation of China (Nos. 12550403, 12174250, 12141404, 12350403, 12174249, 92265102, 12374045), the Innovation Program for Quantum Science and Technology (Nos. 2021ZD0302600 and 2021ZD0302500), the Natural Science Foundation of Shanghai (No. 22ZR1430900). S.J., T.L., X.L. and S.W. acknowledge the Shanghai Jiao Tong University 2030 Initiative Program B (WH510207202). Z.C. acknowledges the support of NSAF (No. U1530402). T.L. and S.J. acknowledge the Yangyang Development Fund. K.W. and T.T. acknowledge support from the JSPS KAKENHI (Grant Numbers 21H05233 and 23H02052), the CREST (JPMJCR24A5), JST and World Premier International Research Center Initiative (WPI), MEXT, Japan. Zhiyuan Future Scholar Program (Grants No. ZIRC2024-02, ZIRC2025-03). Device fabrication is supported by the Micro-nano Fabrication Platform of the School of Physics and Astronomy at Shanghai Jiao Tong University. Y.Z. was supported by Max Planck partner lab on quantum materials. N.M. and C.X.. acknowledge financial support by the Deutsche Forschungsgemeinschaft (DFG, German Research Foundation) through the Würzburg-Dresden Cluster of Excellence ctd.qmat – Complexity, Topology and Dynamics in Quantum Matter (EXC 2147, project-id 390858490).

*These authors contributed equally to this work.
†Contact authors: yangzhang@utk.edu, dqian@sjtu.edu.cn, chenzq@hpstar.ac.cn, swjiang@sjtu.edu.cn

*These authors contributed equally to this work.
†Contact authors: yangzhang@utk.edu, dqian@sjtu.edu.cn, chenzq@hpstar.ac.cn, swjiang@sjtu.edu.cn

*These authors contributed equally to this work.
†Contact authors: yangzhang@utk.edu, dqian@sjtu.edu.cn, chenzq@hpstar.ac.cn, swjiang@sjtu.edu.cn

*These authors contributed equally to this work.
†Contact authors: yangzhang@utk.edu, dqian@sjtu.edu.cn, chenzq@hpstar.ac.cn, swjiang@sjtu.edu.cn